\documentclass[sigconf]{acmart}
\AtBeginDocument{%
  \providecommand\BibTeX{{%
    \normalfont B\kern-0.5em{\scshape i\kern-0.25em b}\kern-0.8em\TeX}}}

\usepackage{multirow}
\usepackage{todonotes}
\setcopyright{acmcopyright}
\copyrightyear{2018}
\acmYear{2018}
\acmDOI{XXXXXXX.XXXXXXX}

\acmConference[SC23]{Make sure to enter the correct
  conference title from your rights confirmation emai}{Nov. 12--17,
  2023}{Denver, CO}
\acmPrice{15.00}
\acmISBN{978-1-4503-XXXX-X/18/06}

\begin{document}

\title{Joint inversion of Time-Lapse Surface Gravity and Seismic Data for Monitoring of 3D CO$_2$ Plumes via Deep Learning}

\author{Adrian Celaya}
\affiliation{%
  \institution{Rice University and TotalEnergies EP Research and Technology USA}
  \city{Houston}
  \state{TX}
  \country{USA}
}

\author{Mauricio Araya-Polo}
\affiliation{%
  \institution{TotalEnergies EP Research and Technology USA}
  \city{Houston}
  \state{TX}
  \country{USA}
}

\renewcommand{\shortauthors}{Celaya and Araya-Polo}

\begin{abstract}
We introduce a fully 3D, deep learning-based approach for the joint inversion of time-lapse surface gravity and seismic data for reconstructing subsurface density and velocity models. The target application of this proposed inversion approach is the prediction of subsurface CO$_2$ plumes as a complementary tool for monitoring CO$_2$ sequestration deployments. Our joint inversion technique outperforms deep learning-based gravity-only and seismic-only inversion models, achieving improved density and velocity reconstruction, accurate segmentation, and higher R-squared coefficients. These results indicate that deep learning-based joint inversion is an effective tool for CO$_2$ storage monitoring. Future work will focus on validating our approach with larger datasets, simulations with other geological storage sites, and ultimately field data.
\end{abstract}





\maketitle

\section{Introduction}
Reducing CO$_2$ concentration in the atmosphere is critical to control climate change. These efforts involve implementing various technologies, such as efficient fossil-based fuel consumption, expanding absorption sources through afforestation/reforestation, and adopting CO$_2$ capture, utilization, and storage (CCUS) techniques.

Among the CCUS technologies currently deployed worldwide, CO$_2$ geological storage has emerged as a promising approach. This technology involves capturing CO$_2$ from fixed sources or directly from air, then injecting it into underground formations. 

Injecting CO$_2$ is just part of the process; during and after injection, ensuring the integrity of these geological sites necessitates ongoing monitoring. Regulatory authorities mandate the demonstration of storage volume containment and the detection of any potential CO$_2$ leakage or unwanted migration. Given the time scales involved in monitoring CO$_2$ storage sites, traditional monitoring techniques based on borehole sensors or surface seismic monitoring may not be practical or economically viable. Remote sensing by other modes, such as gravity, might be economically viable and technically feasible if combined with traditional seismic approaches.

Once data from the field is recorded, geophysical inversion techniques are deployed. These widely used techniques for interpreting data sets and recovering subsurface physical models can be crucial in monitoring CO$_2$ storage sites \cite{fawad2021monitoring, alyousuf2022}. The recovered models, which encompass parameters such as velocity, density, resistivity, or saturation, provide essential information about the structural and compositional characteristics of the subsurface.

Integrating multiple datasets collected over the same area is known as geophysical inversion \cite{hu2023,  joint-inversion}. This approach enhances the reconstruction of subsurface structures and facilitates the identification of changes or anomalies associated with activities like CO$_2$ storage \cite{um2022}. However, effectively integrating the information embedded in multiple geophysical datasets presents a practical challenge from a computational point of view (i.e., storage, memory, and compute time), which compounds the inherent difficulties of solving nonlinear inversion problems \cite{tarantola1982}. These challenges especially become apparent when working with realistic 3D synthetic or field data.

In this work, we develop an effective \textit{supervised} 3D deep learning (DL)-based inversion method to recover high-resolution subsurface CO$_2$ plumes from both surface gravity and seismic data. To the best of our knowledge, this is the first fully 3D approach for the joint inversion of surface gravity and seismic data, which is tested on realistic, physics-simulated CO$_2$ plumes.

\section{Previous Work}
Conventional inversion methods find a model with the minimum possible structure and whose forward response fits the observed data \cite{degrood1990, li1998, nabighian2005}. The minimum structure is achieved by minimizing model roughness through a least squares regression, resulting in a smooth model. While the least squares regression produces smooth subsurface models, the predicted models are often larger and exhibit smaller density values than the actual model \cite{boulanger2001, rezaie2017}. 

DL is an emerging alternative to traditional geophysical inversion \cite{jin2017, araya2018, kim2018, yang2019}. Over the last several years, deep convolutional neural networks (CNNs) have achieved state-of-the-art results in various computer vision applications such as image classification, segmentation, and generation \cite{imagenet, brats2, stylegan}. CNNs have recently been used for inversion of seismic \cite{adler21, chen2020, li2020}, electromagnetic imaging \cite{colombo2020, oh2020}, electrical resistivity \cite{depth-weighting, shahriari2020}, and time-lapse surface gravity data \cite{yang2022, wang2022, huang2021, celaya2023}. However, these works focus solely on inverting a single modality and do not explore joint inversion with multiphysical data. 

Using simulated CO$_2$ plumes from the onshore Kimberlina site, Um et al. developed a 2D DL architecture to perform a joint inversion with seismic, electromagnetic, and gravity data \cite{um2022}. Additionally, they use a modified version of their architecture to invert their imaging data types individually. In each case, their DL-based approach can recover CO$_2$ plumes. However, their approach still does not perform DL-based inversion in a fully 3D setting. Hu et al. present a physics-informed DL-based approach for inverting electromagnetic and seismic data for recovering subsurface anomalies \cite{hu2023}. While their approach successfully reconstructs the anomalies, this approach also uses 2D data and does not consider 3D data or the computational cost of implementing physics-informed inversion for such data; as opposed to our work, which implements fully 3D joint inversion with seismic and surface gravity data.
\section{Problem Statement}
Classical inversion techniques aim to minimize a cost function that measures the difference between the forward response of a given subsurface model and the observed data. Let $F$ be our forward operator, $\mathbf{m}$ be a subsurface model, and $\mathbf{d}_{\text{obs}}$ be the observed data. Then the classical inversion problem can be written as 
\begin{align} \label{inv-single}
    \min_{\mathbf{m}} ||F(\mathbf{m}) - \mathbf{d}_{\text{obs}}||_2^2.
\end{align}
Note that this problem takes a single input (i.e., $\mathbf{d}_{\text{obs}}$) and produces a single predicted subsurface model. 

In contrast, joint inversion is an extension of the classical formulation given by (\ref{inv-single}) that maps multiple inputs (i.e., gravity, seismic, and electromagnetic data) to multiple outputs (i.e., density, velocity, and resistivity models). For a given number of inputs $\mathbf{d}_{\text{obs}}^1, \dots, \mathbf{d}_{\text{obs}}^n$ and appropriate forward operators $F_1, \dots, F_n$, our joint problem is given by
\begin{align} \label{eqn:joint}
    \min_{\mathbf{m}^1, \dots \mathbf{m}^n} \sum_{i=1}^n ||F_i(\mathbf{m}^i) - \mathbf{d}_{\text{obs}}^i||_2^2 + \alpha\sum_{i \neq j} \Phi(\mathbf{m}^i, \mathbf{m}^j),
\end{align}
where $\Phi$ is a coupling function used to link different physical models via known petrophysical relations or other metrics like SSIM, and $\alpha$ is a parameter controlling the contribution of the coupling term \cite{moorkamp2011framework, lelievre2012joint}. Joint inversion is much more time-consuming than independent inversions because of the additional terms in the cost function and the need to exchange information between different models via the coupling terms \cite{hu2023}. There also is a need to determine or adjust the weighting parameter $\alpha$, which adds another layer of complexity to the standard joint inversion method \cite{hu2023}.

Our goal is to train a CNN that can accurately predict the changes in subsurface density and velocity given corresponding variations in surface gravity and seismic data. 

\section{Data Preparation}
Located 60km off the western coast of Norway, the Johansen formation is a promising CO$_2$ storage site with a theoretical capacity exceeding 1Gt of CO$_2$. The formation encompasses an aquifer with an approximate thickness of 100m, spans 100km in the north-south direction and 60km from east to west, and is at a depth that ranges from 2,200m to 3,100m below sea level. This setting provides optimal pressure and temperature conditions for injecting CO$_2$ in a supercritical state. 
  
We generate data based on the Johansen formation using the process described in \cite{celaya2023}. That is, we generate a number of distinct geological realizations that vary in porosity and permeability. We conduct a fluid flow simulation that assumes a 100-year injection period followed by a 400-year migration period. This process produces subsurface changes in density. To convert these density models to $V_p$ models for forward seismic modeling, we use the following conversion:
\begin{align} \label{eqn:rockprop}
    V_p(\mathbf{r}) = \sqrt{\frac{\kappa + \frac{4}{3}\mu}{\rho(\mathbf{r})}},
\end{align}
where $\mathbf{r}$ is the spatial coordinate in our model, $\kappa = 8.14$GPa is the bulk modulus, $\mu = 1.36$GPa is the shear modulus, and $\rho$ is the density value \cite{hoek1981, pariseau2017}. Here, we assume that the formation comprises 80\% shale and 20\% sandstone to get the values for $\kappa$ and $\mu$ \cite{johansen1,eigestad}. Using this process, we generate 180 density/velocity pairs. For preprocessing, we resample each density and velocity model from their original resolution of 440$\times$530$\times$145 to 256$\times$256$\times$128.

\begin{figure}[ht!]
    \centering
    \includegraphics[width=0.95\columnwidth]{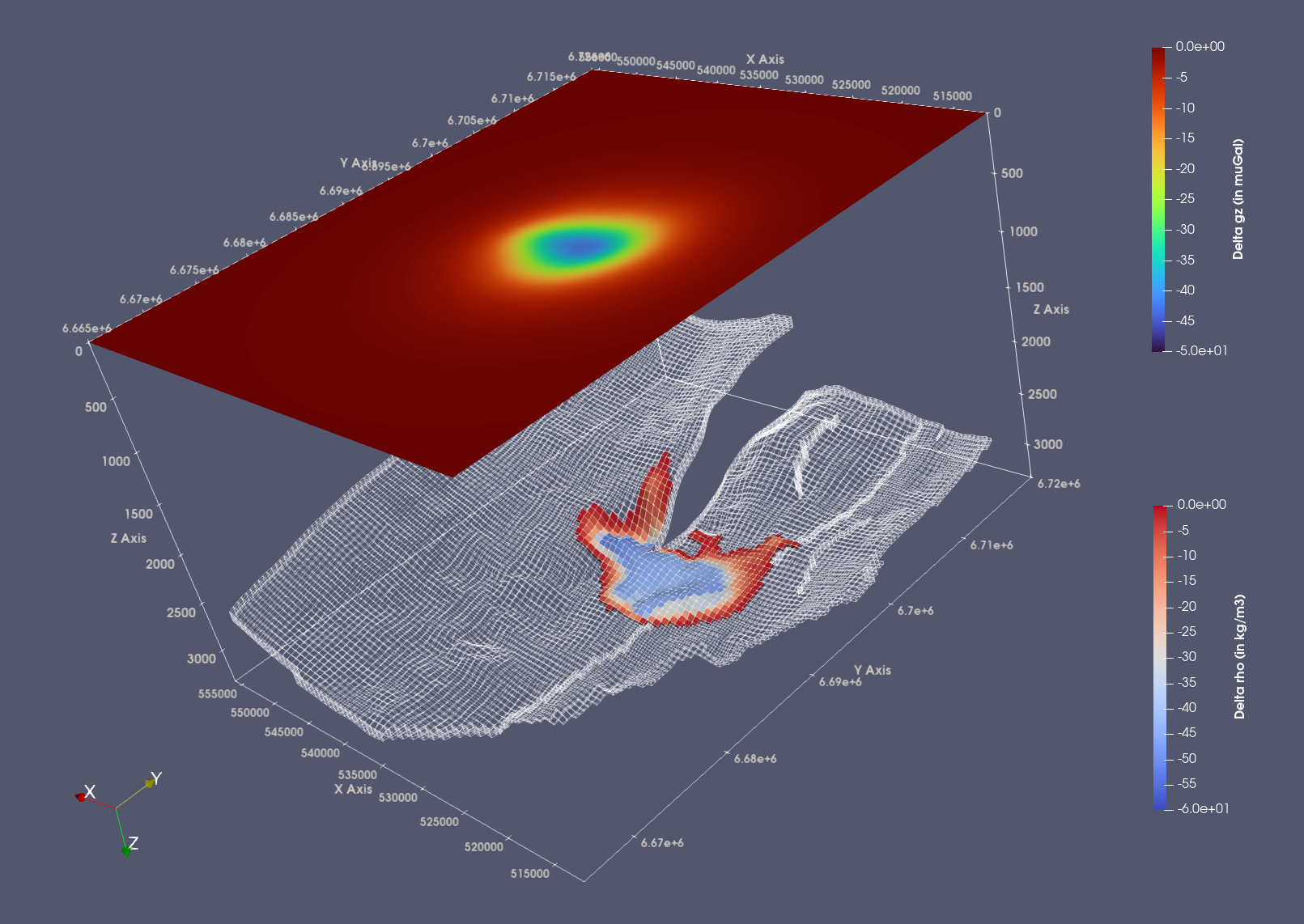}
    \caption{Illustration of a change in surface gravity for a given density perturbation in the subsurface \cite{celaya2023}. \label{io-example}}
\end{figure}
\subsection{Modeling Gravity}
Given a density perturbation $\Delta \rho$ observed between the current and original (i.e., base) acquisition, the gravity field recorded at a station located at $\mathbf{r'}$ can be expressed as:
\begin{align}
	\mathbf{g}(\mathbf{r'})= \gamma \iiint_V \frac{\mathbf{r} - \mathbf{r'}}{|| \mathbf{r} - \mathbf{r'}||_2^3 } \Delta \rho (\mathbf{r}) dV
\end{align}
where $V$ is the volume of the reservoir, and $\gamma$ is Newton's gravitational constant. For more details on gravity modeling, see \cite{celaya2023}. Assuming that gravity sensors are placed in a uniform grid every 500m, our surface gravity maps are size 88$\times$106. An illustration of a change in surface gravity for a given density permutation in the subsurface is shown in Figure \ref{io-example}. 

\subsection{Forward Seismic Modeling}
Forward seismic modeling approximates the behavior of seismic waves propagating through a mechanical medium $\mathbf{m}$ and is given by the elastic wave equation:
\begin{align} \label{elastic}
    \frac{\partial^2 u}{\partial t^2} - \mathbf{V}_p^2 \nabla (\nabla \cdot u) - \mathbf{V}_s^2 \Delta u = \mathbf{f},
\end{align}
where $\mathbf{u} = \mathbf{u}(x, y, z, t)$, is the seismic wave displacement, $\mathbf{V}_p$ is P-wave velocity (compression/rarefaction), $\mathbf{V}_s$ is S-wave velocity (shear stress), and $\mathbf{f}$ is the perturbation source (i.e., shot) function \cite{gelboim2022}. While (\ref{elastic}) more accurately describes seismic wave propagation, it is often preferred (as in this work) to approximate the solution $\mathbf{u}$ by the acoustic wave equation, which assumes only P-waves and requires less computational resources and parameters, as compared to solving (\ref{elastic}) \cite{gelboim2022}.

3D seismic data consumes a large amount of memory and storage, up to dozens of TB for the raw data in our case. We compute spatial decimation evenly, but temporal, which in seismic signals represents depth, decimation favors samples that convey information in the area of interest (i.e., the reservoir). Further, we collapse our seismic data by adding the recorded data from each shot and then boost the later time signals by multiplying by a monotonically increasing function. This method is fully described in \cite{gelboim2022}. Finally, we resize each collapsed and boosted seismic cube from its original resolution of 167$\times$154$\times$941 to 256$\times$256$\times$512 by first taking every other slice (i.e., time step) in the z-direction and then linearly interpolating to the final resolution.

\begin{figure}[ht!]
    \centering
    \includegraphics[width=0.7\columnwidth]{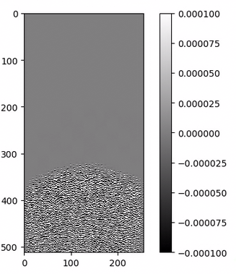}
    \caption{Example of seismic difference cube. \label{io-example-seismic}}
\end{figure}

Like forward gravity modeling, we look at the differences between the original or base seismic data and the current data. Figure \ref{io-example-seismic} illustrates an example of the change in seismic data from the original and current acquisitions. 

\section{Methods}
\label{sec:methods}
\subsection{Network Architecture}
We use a modified 3D U-Net architecture to map 2D surface gravity maps and 3D seismic cubes to subsurface changes in density and velocity. Each input is fed into a dedicated encoder branch. The first step for the surface gravity encoder branch is to resize the input to match the reservoir geometry via linear interpolation. Then the 2D features are converted into a 3D volume using a pointwise convolution, where the number of channels equals the subsurface model's depth (i.e., the output). This resulting 3D volume is the input to the first 3D encoder branch of the network. The seismic encoder branch starts with a 3D seismic cube as input. However, the seismic cubes are larger in the depth dimension than the network output. To address this, we use two convolutional layers with strides 1$\times$1$\times$2 to reduce the depth dimension of the seismic cube from 512 to 128. The resulting resized seismic features are the input to the second encoder branch of our architecture. In each resolution level of our encoder branch, we apply two convolutional layers and downsampling via max pooling four times. A convolutional layer consists of two convolutions, each followed by batch normalization and a ReLU non-linearity. 

At the bottleneck of our architecture, we concatenate the encoded seismic and gravity features and apply two convolutional layers to merge the individual features. We also apply autoencoder regularization. However, because we have two inputs, we separately decode the seismic and gravity features (before concatenation and convolution) to reconstruct their respective inputs.

\begin{figure*}[ht!]
    \centering
    \includegraphics[width=0.95\textwidth]{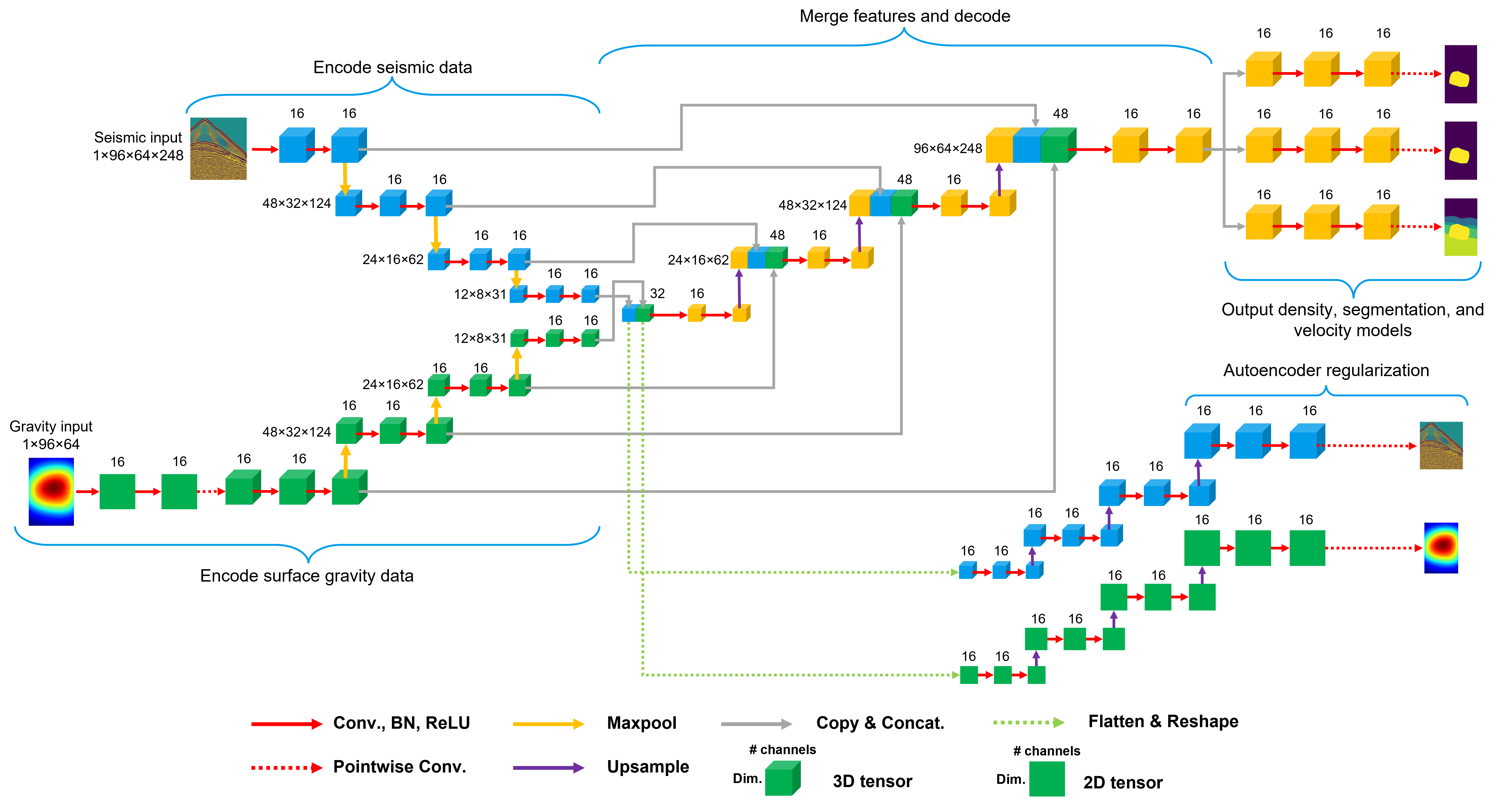}
    \caption{Detailed sketch of our proposed architecture for jointly inverting surface gravity and seismic data. Our architecture uses two separate encoder branches for each input and decodes them jointly. The output of the decoder branch splits into three separate outputs: two regression branches to predict density and velocity and one for segmenting the plume. Additionally, this architecture uses autoencoder regularization. \label{architecture}}
\end{figure*}

Our decoder branch upsamples the combined features and concatenates them with the corresponding features from the encoder branches in the skip connections. Like with the network proposed by \cite{celaya2023}, we split the output of the decoder branch into three separate outputs: two regression branches to predict density and velocity and one for segmenting the plume. Figure \ref{architecture} shows a detailed sketch of our proposed architecture.

\subsection{Loss Function}
Our loss function consists of three components - segmentation, regression, autoencoder losses. 

The segmentation loss is the Generalized Dice Loss (GDL) proposed by Sudre et al. \cite{gdl}. Unlike the original Dice loss proposed by \cite{vnet}, the GDL uses weighting terms to account for class imbalance. This loss function is given by 
\begin{align}
    \mathcal{L}_{gdl} = \frac{\sum_{k=1}^C w_k ||T^k - P^k||_2^2}{\sum_{k=1}^C w_k \left(||T^k||_2^2 + ||P^k||_2^2\right)},
\end{align}
where $C$ denotes the number of segmentation classes, $P^k$ is the $k$-th class in the predicted mask, and $T^k$ is the same for the ground truth. The term $w_k$ is the weighting term for the $k^{th}$ class and is given by $w_k = \left(\frac{1}{\sum_{j=1}^{C} \frac{1}{N_j}}\right) \frac{1}{N_k}$. Here, $N_k$ is the total number of pixels belonging to the class $k$ over the entire dataset. Note that the weights $w_k$ are pre-computed and remain constant throughout training. In our case, the number of segmentation classes equals two; background and foreground. The computed class weights are approximately 0.0015 and 0.9985 for the background and foreground classes, respectively.

The regression and autoencoder losses use the mean squared error loss for their respective inputs. For a general input pair $(T, P)$, this loss is given by $\frac{1}{N} ||T - P||_2^2$. For the regression tasks (i.e., density and velocity reconstruction) the inputs to this loss function are the true and predicted density and velocity models. For the autoencoder branches, the inputs to this loss function are the true and reconstructed gravity and seismic inputs.

Our overall loss function is a weighted convex combination of the previously described components and is given by
\begin{align}
    \mathcal{L} = 0.375\mathcal{L}_{\rho} + 0.2\mathcal{L}_{gdl} + 0.375\mathcal{L}_{v} + \frac{0.05}{2}\left( \mathcal{L}_{ae}^{grav} + \mathcal{L}_{ae}^{seis}\right),
\end{align}
where $\mathcal{L}_{\rho}$, $\mathcal{L}_{gdl}$, $\mathcal{L}_{v}$, $\mathcal{L}_{ae}^{grav}$, and $\mathcal{L}_{ae}^{seis}$ are the density model, segmentation, velocity model, gravity autoencoder, and seismic autoencoder losses, respectively. Note that we select these weights via a partial grid search.

\subsection{Training and Testing Protocols}
To train our neural networks, we use the Adam optimizer \cite{adam} with an initial learning rate of 0.001 and a cosine decay schedule with restarts \cite{cosine-lr}. We train our model to convergence ($\approx 400$ epochs) with a batch size of 8. We use 80\% of our dataset as a training set, use 10\% of the training data as a validation set, and use the remaining 20\% of the data as a test set. To compare the effect of joint inversion vs. inversion with only gravity or seismic data, we use the architecture described in \cite{celaya2023} for the portions of our dataset corresponding to either gravity or seismic inversion. Note that the models used for individual inversion only output the properties corresponding to that particular inversion problem. For example, the network used for purely seismic inversion only produces a velocity model as a prediction.

To evaluate the validity of our predicted inversions, we utilize the following metrics: mean squared error in kg/m$^3$ between the true and predicted density models, mean squared error in m/s between the true and predicted velocity models, mean squared error in $\mu$Gals between the observed data and the gravity response of the predicted density model, the R-squared coefficient between the true and predicted models (for density and velocity), and the Dice coefficient between the non-zero masks of the true and predicted plume geometry. 

Our models are implemented in Python using PyTorch (v2.0.1) and trained on four NVIDIA A100 GPUs \cite{pytorch}. At test time, our DL-based methods produce predictions of size 256$\times$256$\times$128. We resample this output to the original grid resolution of 440$\times$530$\times$145 via linear interpolation to produce our final prediction. Given the sample size, we develop a data parallelism approach by using PyTorch's Distributed Data Parallel implementation. The in-node scalability is nearly ideal, with epochs taking 190 seconds running on 1 GPU and ending up in 45 seconds when running on 4 GPUs.  

\begin{table*}[!ht]
\begin{tabular}{lcccccc}
\hline
\multicolumn{1}{c}{\multirow{2}{*}{Method}} & \multirow{2}{*}{MSE (kg/m$^3$)} & \multirow{2}{*}{MSE (m/s)} & \multirow{2}{*}{MSE ($\mu$Gal)} & \multirow{2}{*}{Dice} & \multicolumn{2}{c}{R-Squared} \\ \cline{6-7} 
\multicolumn{1}{c}{} &  &  &  &  & Density & Velocity \\ \hline
Gravity & 0.311 (0.215) & - & \textbf{0.601 (1.185)} & 0.589 (0.034) & 0.769 (0.12) & - \\
Seismic & - & 0.121 (0.100) & - & 0.579 (0.096) & - & 0.715 (0.175) \\
Joint & \textbf{0.268 (0.174)} & \textbf{0.084 (0.057)} & 0.634 (1.241) & \textbf{0.621 (0.038)} & \textbf{0.801 (0.095)} & \textbf{0.800 (0.095)} \\ \hline
\end{tabular}
\caption{Joint inversion vs. inversion with only gravity or seismic data. Here, we see that our proposed joint inversion generally outperforms gravity and seismic-only inversion for all metrics except for the mean squared error between the observed data and the gravity response of the predicted density model, where the results vs. the gravity-only model are comparable. \label{tab:results}}
\end{table*}

\begin{figure}[!ht]
    \centering
    \includegraphics[width=\columnwidth, height=1.5in]{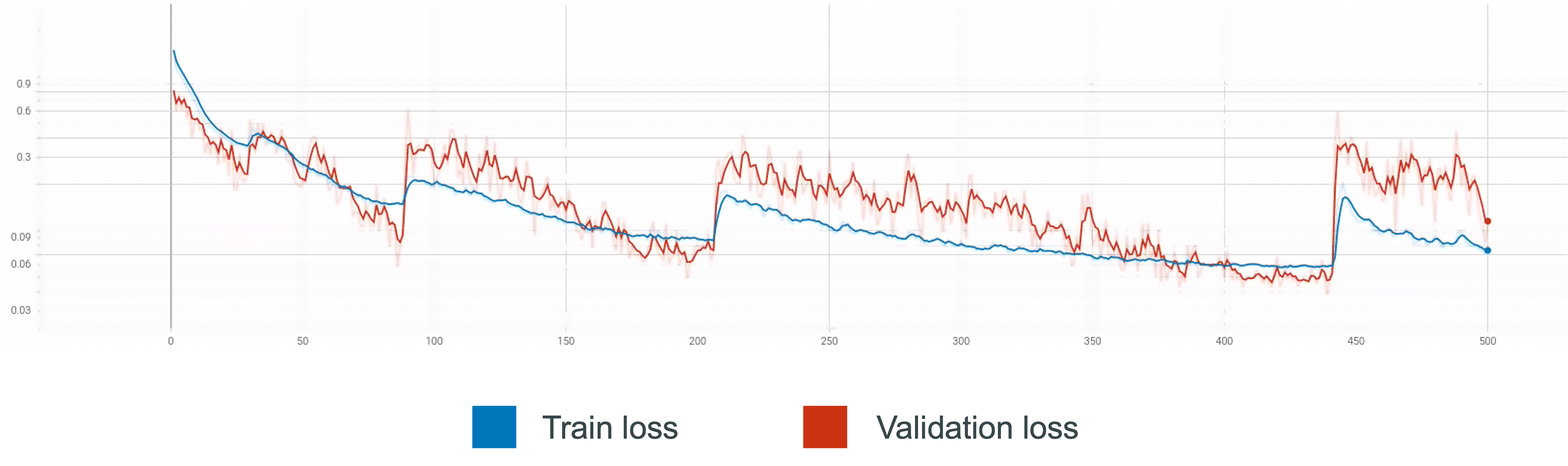}
    \caption{Training and validation loss curves on a logarithmic scale for our DL-based joint inversion. Here, both losses converge, indicating that our proposed joint inversion architecture successfully learns a mapping from our surface gravity and seismic data to 3D subsurface density and velocity models. \label{loss}}
\end{figure}
\section{Results}
We train our joint inversion model using the methods described in Section \ref{sec:methods}. Figure \ref{loss} shows the training and validation loss curves on a logarithmic scale. This figure shows that both losses converge, indicating that our proposed joint inversion architecture successfully learns a mapping from our surface gravity and seismic data to 3D subsurface density and velocity models. 

\begin{figure*}[ht!]
    \centering
    \includegraphics[width=0.95\textwidth, height=3in]{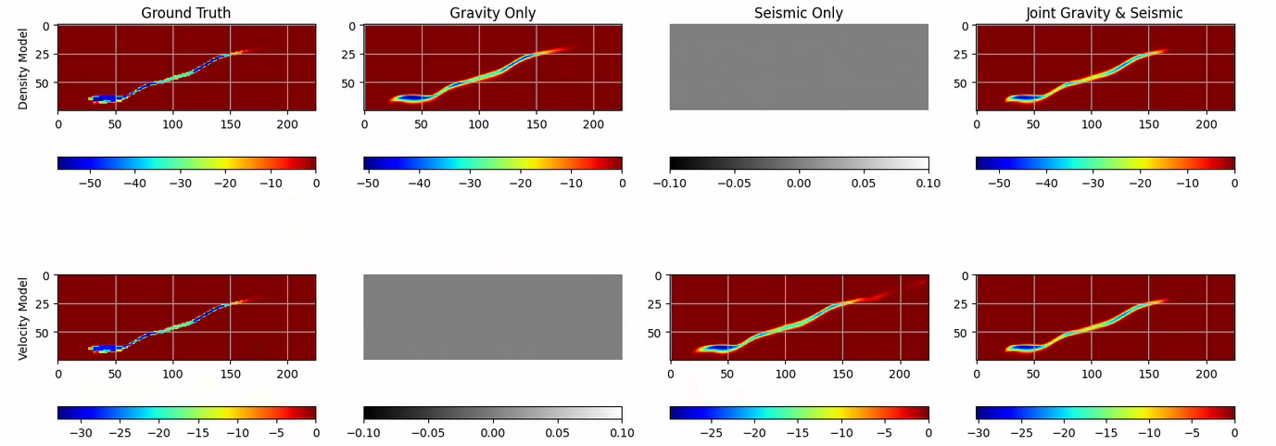}
    \caption{From left to right 2D cross-section slice from ground truth, gravity-only, seismic-only, and joint models. The top row shows density models, and the bottom row shows velocity models. Visually, the density models produced by the gravity-only and our joint architectures are similar. However, visual differences exist between the seismic-only reconstructed velocity model and the velocity model produced from our joint architecture (i.e., top right corner). \label{preds}}
\end{figure*}

\begin{figure}[ht!]
    \centering
    \includegraphics[width=0.95\columnwidth, height=1.8in]{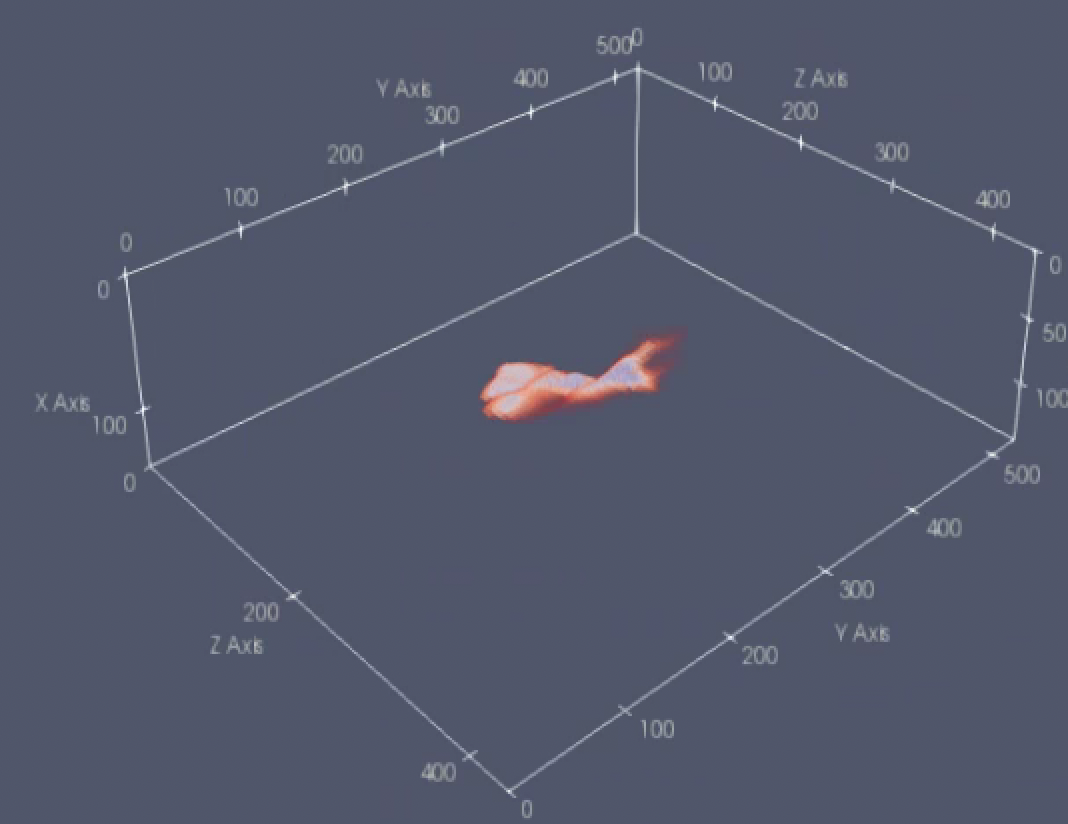}
\includegraphics[width=0.95\columnwidth, height=1.8in]{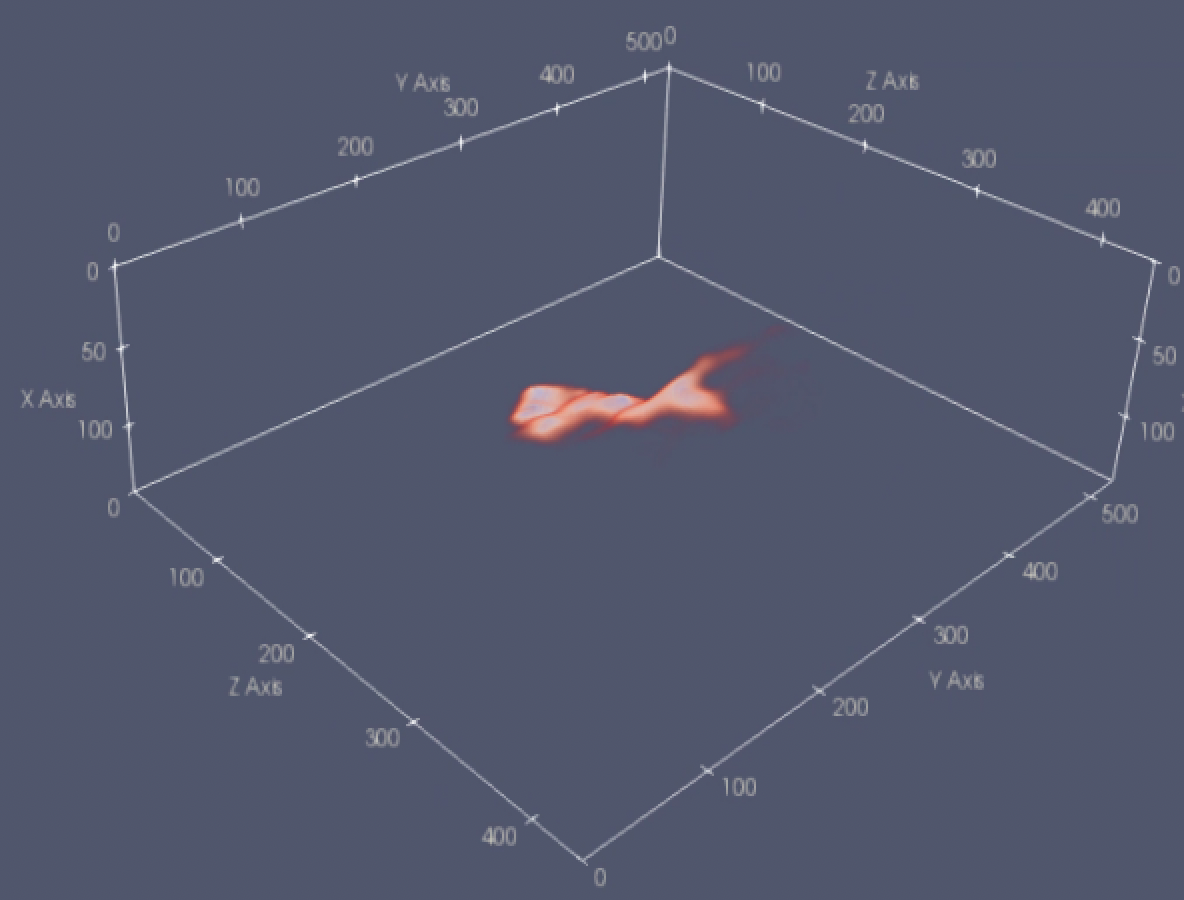}
    \caption{3D view of label (top) -prediction (bottom) pair of velocity variation given the CO$_2$ plume location.\label{velocity_pred}}
    \end{figure}

In Table \ref{tab:results}, we see that our proposed joint inversion generally outperforms gravity and seismic-only inversion for all metrics except for the mean squared error between the observed data and the gravity response of the predicted density model, where the results vs. the gravity only model are comparable. Figure \ref{preds} shows a 2D cross-section slice from predictions from each model. Visually, the density models produced by the gravity-only and our joint architectures are similar. However, visual differences exist between the seismic-only reconstructed velocity model and the velocity model produced from our joint architecture (i.e., top right corner); those can also be observed in Figure~\ref{velocity_pred} for a different sample (i.e., at different times of plume's migration).

\section{Discussion}
Our results demonstrate the potential benefits of DL-based joint inversion. The joint inversion model consistently outperforms DL-based gravity-only and seismic-only inversion models across various evaluation metrics. This performance suggests that the fusion of surface gravity and seismic data can lead to more accurate subsurface models. The improved performance of the joint inversion model in terms of density and velocity reconstruction, segmentation accuracy, and R-squared coefficients indicates the effectiveness of the proposed approach in capturing the complexity of subsurface CO$_2$ plumes.

The benefits of DL-based joint inversion are limited from a computational perspective because combining two different datasets requires more memory and time during training. Our joint inversion model takes roughly 45s per epoch on 4 A100 GPUs with a batch size of 8. In contrast, DL-based gravity inversion takes roughly 20s per epoch, and DL-based seismic inversion takes roughly 30s per epoch for the same number of GPUs and batch size. Additionally, joint inversion takes just over 400 epochs to converge to a solution vs. 200 for both the gravity and seismic-only approaches. The greater number of epochs is possibly explained by the more complex relationship our joint inversion approach has to resolve vs. the single mode methods. Regarding inference, our joint model is comparable to the gravity and seismic-only models, producing predictions in less than one second on a single A100 GPU.

We utilize the PocketNet approach proposed by \cite{pocketnet} in our joint architecture. This approach takes advantage of the similarity between the U-Net architecture and geometric multigrid methods to drastically reduce the number of parameters \cite{pocketnet, mgnet}. Additionally, we replace the transposed convolution with trilinear upsampling. With these modifications, we reduce the number of parameters from roughly 33,000,000 to 349,000, yielding a roughly 40\% decrease in the time per epoch. Additionally, these modifications allow us to use a larger batch size (8 instead of 4). 

While developing our joint inversion model, we observe that the optimization landscape is tricky, with some local minimums corresponding to good values for density model misfit and R-squared and vice versa for velocity models. Adding a coupling term like in the classical joint inversion formulation given by \ref{eqn:joint} may help avoid getting trapped in local minimums during training. Future work will focus on formulating and testing a term based on the Dice score. The intuition here is to enforce via our modified loss function that the plumes occupy the same physical space.

Our data comes from simulations of a single geologic formation (Johansen). Further work is needed to test the generalization capability of our joint inversion model to other datasets based on simulations of other CO$_2$ storage sites (i.e., Snohvit and Kimberlina \cite{alumbaugh2023kimberlina}) and on field data. However, time-lapse CCUS monitoring with gravity (and other non-seismic methods) is a data-poor field. Sufficiently detailed reservoir models for CCUS, especially field data, are hard to come by, and access is limited \cite{celaya2023, alumbaugh2023kimberlina}.

\section{Conclusions}
We developed an effective DL-based joint inversion method to recover high-resolution, subsurface CO$_2$ density and velocity models from surface gravity and seismic data. We train our joint DL architecture on realistic, physics-simulated CO$_2$ plumes, surface gravity, and seismic data. This training approach mirrors real-world site data collection. Our joint inversion outperforms gravity and seismic-only inversion techniques for our selected metrics. While there is room for improvement, the results presented here are promising and represent, to the best of our knowledge, the first fully 3D DL-based joint inversion of surface gravity and seismic data derived from a physics simulation of a proposed CO$_2$ storage site. 


\begin{acks}
This work was supported by TotalEnergies EP Research \& Technology USA, LLC.
\end{acks}

\bibliographystyle{ACM-Reference-Format}
\bibliography{references}










\end{document}